\documentclass[preprint,prc,aps,nofootinbib]{revtex4-1}
\newcommand{\be}{\begin{eqnarray}}
\newcommand{\ee}{\end{eqnarray}}

\newcommand{\order}[1]{ \mathcal{O} \left( #1 \right) }

\newcommand{\ave}[1]{\left\langle #1 \right\rangle}

\newcommand{\eqcomma}{\phantom{A},\phantom{A}}
\newcommand{\lnz}{\ln \mathcal{Z}}
\newcommand{\lqcd}{\Lambda_{\mathrm{QCD}}}
\usepackage{epsfig}
\begin{document}
\title{ Swimming and swirling colorful ghosts:\\ The ideal hydrodynamic limit and non-Abelian gauge symmetries}
\author{Giorgio Torrieri}
\affiliation{IFGW, State University of Campinas, Brazil}
\email{torrieri@ifi.unicamp.br}
 \begin{abstract}
   We show that the ideal fluid local equilibrium limit, defined as the existence of a flow frame $u_\mu$ which characterises the direction of both a conserved entropy current and conserved charge currents is incompatible with non-Abelian gauge theory if local color charge density is non-zero.   Instead, the equation of state becomes dependent on $u_\mu$ via modes which are roughly equivalent to ghost modes in the hydrodynamic limit.  These modes can be physically imagined as a field of ''purcell swimmers'' whose ''arms and legs'' are outstretched in Gauge space.  Also, vorticity should couple to the Wilson loop  via the chromo-electro-magnetic field tensor, which in local equilibrium is not a ''force'' but instead represents the polarization tensor of the gluons. We show that because of this coupling vorticity also acquires swirling non-hydrodynamic modes.
We then argue that these swirling and swimming non-hydrodynamic modes are the manifestation of gauge redundancy within local equilibrium, and speculate on their role in quark-gluon plasma thermalization
 \end{abstract}
 \maketitle
 \section{Introduction}
 \subsection{The ideal fluid and color charge density}
An ideal fluid, at its most fundamental level, can be thought of as a system where every ``small cell'' is close to its local equilibrium state, defined by local entropy maximization \cite{kodama,spal} subject to the constraints inherent in the symmetries of the microscopic theory.
Lagrangian techniques can be used to develop an effective theory just out of this assumption \cite{nicolis1,shift}.   In the past, this approach has been extended to include approximately equilibrated systems \cite{gtlagrangian} and systems with microscopic polarization \cite{gt1,gt2,gt3}.   A logical extension, relevant to quark gluon plasma \cite{kodama}, would be to include the full Non-Abelian gauge symmetries of microscopic QCD. 

This has as far not been done, as what was intended as ``Lagrangian non-Abelian fluid dynamics'' previously \cite{jackiw} meant something quite different.
These works, as well as any others \cite{wong,heinz,surowka,mrow,whitening}  developed via an extension of a Vlasov-type equation to a Non-Abelian theory, or a charged ideal fluid coupled to a Yang-Mills field, a solution to classical Yang-Mills equations.  These approaches presuppose a mixture between a thermalized high-entropy fluid, with small mean free path, and a coherent field, which carries zero entropy and is characterized an infinite mean free path.
 It is not clear weather this system is amenable to a good effective theory expansion, since there are quite a few very different length-scales in the problem.   It is therefore not surprising these approaches often lead to instabilities, which have also been argued to induce a rapid effective equilibration \cite{mrow,asakawa,strickland}.   These instabilities have been argued to lead to a local ''hydronomization without local equilibration'': A rapid isotropization of degrees of freedom means that the system, {\em on average}, looks like a locally equilibrated ideal fluid whose mean free path is small even if the detailed dynamics of the system is more like that of a Vlasov equation.   Hence, ''color hydrodynamics'' would be a good model not for hydrodynamics evolution but for pre-hydrodynamic thermalization.

Recent phenomenology, however, shows that the first approach, assuming that the thermalization timescale is shorter than the color coherence one, is worthy of investigation.
  Fluctuations of elliptic flow \cite{mrowfluct,vogel,kodama}, today investigated via cumulant measurements, 
were suggested to distinguish between real and fake equilibration.   Intriguingly, equilibration seems to be ``real'' (i.e. local within each fluid cell) even, as in pp collisions \cite{cms}, it would require hydrodynamics to work well at sub-hadronic scales.
This would naively require for the fluid to possess thermalized local color charge, and associated Yang-Mills fields which are not just chaotically fluctuating but determined entirely by entropy maximization under the constraints of the local charge and angular momentum balance, in analogy to fluids with a local polarization density \cite{gt1,gt2,gt3}.  No hydrodynamic theory for such a system has to our knowledge as yet been constructed.

Before attempting to construct it, let us motivate its need a bit further.
The relevance of color neutrality to hydrodynamic evolution can be summarized within three scales, dimension of length.   We shall give, inside the $\{...\}$ brackets the current theoretical uncertainty range for each
 \begin{equation}
   \label{scales}
  L_{color}^{-1} \sim \left\{  Q_s-\lqcd \right\}  \eqcomma L_{mfp}^{-1} \sim \{\ \lambda^2 T - 4 \pi T - \tau_0^{-1} \sim Q_s \} \eqcomma L_{debye}^{-1} \sim \{ \sqrt{\lambda} T - 3.4\pi T \}
 \end{equation}
where, respectively, $Q_s$ is the saturation momentum scale, $\tau_0$ the fluid formation scale, $\lqcd$ the QCD non-perturbative scale, $T$ the temperature, and $\lambda$ the 't 'Hooft coupling constant.

Note that $L_{mfp}$ could be either the strong or weak coupling mean free path, or it could refer to the possibly different or possibly comparable formation time of the hydrodynamic phase.  There is some confusion weather $L_{debye}$ depends on the coupling constant, it depends on how it is constructed \cite{adsreview,karch}. But it is clear that $L_{mfp}$ could be $\gg L_{debye}$ or $\sim L_{debye}$.

What about $L_{color}$, the size of the domain of the orientation in color space?  In an AdS/CFT approach, it is difficult to see how the planar limit can avoid for it to be parametrically smaller than the above two scales, but as we know $N=3 \ll \infty$.   
A cursory glance at Baryon lattice configurations \cite{lin} shows that at zero temperature color domains are of roughly baryonic size in configuration space.
A highly boosted baryon's gluon wavefunctions will be modified by a diffusion in rapidity, and, according to popular color glass models the characteristic size of color domains is $Q_s^{-1}$, with corrections of $\order{0.5-2}$ \cite{schenke,domains}.   If, however, the semiclassical regime is not reached, it is not unreasonable that suppose the transverse size of color domains is of the order of 0.5-0.7 the baryonic transverse size.

Hence, in  the possibility that in the initial state $L_{color} \sim L_{debye} \gg L_{mfp}$ is not ruled out by the data.  In fact, if one takes seriously the claim\cite{cms} that the scaling of $v_2$ with cumulant number in $pp$ collisions indicates as good a thermalization as in $AA$ collisions, and uses Bjorken formula extrapolations 
 \begin{equation}
 \order{0.1}\tau_0 R^2 T^3 \sim \frac{dN}{dy} \eqcomma R \sim \order{1}\mathrm{fm} \eqcomma \frac{dN}{dy} \sim \order{10-50} \eqcomma Q_s \sim \order{1} GeV
 \end{equation}
such a hierarchy appears {\em favored}.

In a transport regime, such a situation would be irrelevant since color scrambling of the domains would quickly follow \cite{whitening} on a scale $\order{0.1}L_{mfp}$, but if $L_{mfp} \leq (4 \pi T)^{-1}$ (which is likely if the Knudsen number for the pp system $\ll 1$) and color diffusion is of the order of the mean free path, microscopic scrambing would not occur.
What happens in a non-perturbative regime is far less clear:
Qualitative gedankenexperiments involving such QGP at large scales leads to seemingly contradictory conclusions such as ``orphan quarks`` \cite{orphan} and outright paradoxes \cite{paradox}, while the statistical mechanics of high temperature pQCD contains infrared singularities \cite{linde} which lead to unexpected coupling constant dependencies even in the region where the coupling constant is ``small`` \cite{arnold}.  There are also good theoretical reasons to think that confinement remains across such large scales at any temperature \cite{alkofer}.

However, no quantitative non-perturbative mechanism of dynamical color neutralization across large scales in a hot theory has so far been proposed (we shall remark about positivity violation at the end \cite{alkofer}).   Thus, the idea that initial stages of hydrodynamic evolution are characterized by a non-zero color charge density is at least possible, and the question of how such a regime evolves needs to be tackled.
 
 One can of course assume that a total thermalization of fluid and field degrees of freedom can be well described by a parton cascade, which can be made compatible with non-abelian gauge symmetry \cite{cascade}, leading to the scrambling mechanism described in \cite{whitening}.  However, it is not clear that in the strongly coupled limit the hyerarchy of correlations (the quantum generalization of the BBGKY hyerarchy) can be effectively truncated so only microscopic averages of the parton distribution functions are relevant.  
 For example, in a non-Abelian gauge theory, even in the absence of Fermions, gauge bosons carry spin and its interaction, in addition to simple scattering, contains spin-orbit and spin-spin interactions.   A gas of gauge bosons in equilibrium with non-zero angular momentum will experience non-zero spin alignment for any finite temperature \cite{becfl}.
Hence, hydrodynamics, if a fluid is inviscid enough to carry sizeable vorticity,  self-consistently must include polarization, which is problematic to describe using a Boltzmann or a Vlasov equation, or, for that matter, a theory just in terms of conservation laws and isotropization.  As is shown in \cite{gt1,gt2,gt3}, this can be gotten around by elevating local equilibrium (local maximization of entropy in each cell) as a fundamental principle and building an action principle from local equilibrium.
 
 On the other hand, the whole point of gauge symmetry is the freedom to exchange spatial angular momentum for the longitudinal polarization of the Gauge boson.  For abelian theory, this exchange is ''harmless'' \cite{araujo,goldman} since the mean free path of photons is infinite.
 However, if there are interactions and local equilibrium, since spatial angular momentum is carried by vorticity, this ambiguity seems to contradict the very definition of local thermalization, since polarization is a microscopic density and vorticity is part of flow, a macroscopic variable.  In the following subsection we shall show that this ambiguity is inherent in the definition of local thermalization within relativity
\subsection{Local equilibrium, Gauge symmetry and the energy momentum tensor \label{tmunu}}
In the absence of locally conserved charges, the only conserved current is the energy-momentum tensor $T_{\mu \nu}$ defined canonically as
\begin{equation}
\label{tmunucan}
  T^\nu_\mu \underbrace{=}_{\forall \phi_A} L \delta_\mu^\nu -  \sum_A \left(\frac{\partial L}{\partial (\partial_\nu \phi^A)}\right)\partial_\mu \phi_A     \underbrace{=}_{\phi_A \rightarrow A_\mu} L \delta_\mu^\nu -  \left(\frac{\partial L}{\partial (\partial_\nu A^\alpha)}\right)\partial_\mu A^\alpha 
\end{equation}
where the first equality refers to arbitrary arrays of fields and the second to Lorentz covariant 4-vectors.

In the ideal fluid dynamics limit, local isotropy means that there is a velocity frame field $u^\mu$ at rest with which the system is isotropic, homogeneous, and locally equilibrated.   That fixes the energy-momentum tensor and any internal conserved charge current $J_\mu$ to the form
\begin{equation}
  \label{hydrot}
  T_{\mu \nu} = u_\mu u_\nu (p+e) - p g_{\mu \nu} \eqcomma J_\mu = n u_\mu
\end{equation}
where $p,e,n$ are scalars representing the pressure, energy density and conserved charge density in the co-moving frame.   Together with the local equilibrium condition, which expresses $e,p,n$ in terms of the partition function $\lnz$, the free energy divided by the temperature
\begin{equation}
  \label{lnzdef}
  \exists \lnz \eqcomma  p = T \lnz \eqcomma e=\frac{d\lnz}{d(1/T)} \eqcomma n = T \frac{d\lnz}{d(\mu)}
\end{equation}
these equations will be closed, i.e. solvable from initial conditions.
These equations define the coarse-grained variables and establish a zero-order term.   A gradient expansion can then be developed \cite{kodama,spal} in terms of gradients of $u^\mu,p$ and $n$ multiplied by a microscopic scale (the mean free path or the diffusion lenght), often called the Knudsen number expansion.

The problem is that Eq. \ref{tmunucan}
will typically be Gauge-dependent \cite{hehl,pseudo}, and hence it is not unambiguously defined.  The ambiguity can be expressed as the addition of a total derivative of an anti-symmetric function $\Phi_{\alpha,\beta \gamma}$ (total derivativesof course do not change conservation laws)
 \begin{equation}
   \label{pseudogauge}
  T_{\mu \nu} \rightarrow T_{\mu \nu} + \frac{1}{2} \partial^\lambda
  \left( \Phi_{\lambda,\mu \nu} - \Phi_{\mu,\lambda \nu} - \Phi_{\nu,\lambda \mu} \right)
 \end{equation} 
 A particular choice of $ \Phi_{\nu,\lambda \mu}$ (the exact form is also Gauge dependent) will remove the anti-symmetric part \cite{jackson} and put this tensor as equal to the Gauge-independent object derived from the action $S$
 \begin{equation}  T^{\mu \nu}_B = 2\left. \frac{\delta}{\delta g_{\mu \nu}} S\right|_{g_{\mu \nu}=\mathrm{diag}[1,-1,-1,-1]}
\label{blderiv}
   \end{equation}
 (S is, in an EFT, the logarithm of a partition function $\lnz$).
This new tensor (usually called Belinfante-Rosenfeld tensor) is symmetric, and contains no information regarding how much angular momentum is contained in polarization.   For this reason, works such as \cite{pseudo} have argued that local equilibrium only makes sense when the energy-momentum tensor is defined canonically, with an anti-symmetric component.   This, however, leaves us in an uncomfortable position as it implies local equilibrium itself is ambiguous, since it depends on a Gauge-dependent constraint on the microscopic fields.

   One can see that this issue is more general than transport if  one considers that, as illustrated in
\cite{brauner}, the difference between the different definitions of $T_{\mu \nu}$ boils down to a field redefinition combined with a coordinate transformation, a generalization of the way the Noether angular momentum current is formed.   For a gauge theory, the field redefinition is a Gauge transformation.  It is far from clear how a gauge-violating effective theory can emerge from a microscopically Gauge-invariant dynamics, weather in a weakly or a strongly-coupled regime, since no equivalent of the Higgs mechanism is evident in a gauge theory at local equilibrium \cite{elitzur}.

In the lagrangian picture, local equilibrium is equivalent to the KMS condition of time periodicity \cite{kubo} in the co-moving frame, which implies that flow is a Killing vector of the local Lagrangian coordinates \cite{nicolis1,shift}.   Hence, if the different definitions of $T_{\mu \nu}$ are to be equivalent, as an alternative to Gauge symmetry being violated, the dynamics of a locally thermalized system with non-zero charge potential might not be determined by locally co-moving variables alone when the color charge density is non-zero.

Physically, if one thinks about what non-Abelian gauge symmetry entails, this would not be so surprising.
Invariance of the dynamics to the total derivative terms is a reflection of the fact that quantities such as entropy, charge and, for the non-gravitational dynamics of relevance here, energy density, are only physical up to a ''zero point''.   This zero point is gauge dependent, but physics only depends on gradients of these quantities.   In a sense, gauge symmetry adds as a zero point the quantity of the longitudinal spin v.s. longitudinal angular momentum present in the system.    But, and this is a crucial point w.r.t. this paper, non-Abelian gauge theories have ''pure gauge'' terms with arbitrarily large and complex gradients. If entropy gradients are locally gauge dependent, a definition of a flow field $u_\mu$ w.r.t. the system is in local equilibrium becomes problematic   \footnote{A globally equilibrated color-neutral state in the Grand Canonical ensemble is of course well-defined and understood using lattice techniques.  The relation between such a system and the locally equilibrated ideal fluid examined here is the same as the relation between a hydrostatic bath in global equilibrium, and the same bath with sound-waves bouncing around it. Even without Gauge theories the relationship between the two setups can be extraordinarily subtle, something the last section will discuss further}.

 In this work we shall confirm that the reasoning above is indeed correct, by treating hydrodynamics as a ''bottom-up'' effective theory, where the hydrodynamic limit is defined not in terms of an underlying theory but in terms of its symmetries and explicitly determining the lowest derivative field configurations on which the free energy depends.   Since in the Lagrangian picture the free energy determines the action, our demonstration should be valid in any system close to local equilibrium.

The definition of hydrodynamics in Eq. \ref{hydrot} hides in it an assumption whose validity in a Non-Abelian gauge theory is dubious \cite{gt0}:  What Eq. \ref{lnzdef} gives is not the guaranteed configuration, but the most likely one, the one with most microstates (each of which is equally likely).   Further derivatives of $\lnz$ will give fluctuations, which need to be small in each coarse-grained cell to avoid stochastic ``kicks'' in the fluid (note that the planar limit is enough to eliminate such kicks, as they are of $\order{1}$ in color counting).

Even more crucially fluctuations driven by microstates within different cells must be either uncorrelated, or at most correlated according to the same Gradient expansion as the Knudsen number \cite{kodama,spal}.  Entanglement between quantum states of two fluid cells whose amount is not proportional to flow separation, for example, would impair the hydrodynamic limit.
It is this assumption that in principle could fail within a non-Abelian gauge theory, provided the fluid has a non-zero net color charge density.  It is well-known that Gribov ambiguities arise within a scale that can, for non-conformal theories such as QCD translate to a spacetime scale \cite{marcelo0,marcelo}.  For a locally but not globally equilibrated theory, this could translate into a degeneracy of entropy maxima correlating different cells. This degeneracy has a gradient which has nothing to do with the flow and temperature gradients, and is rather related to the gradient of extended equilibrium objects conjectured to exist in QCD (``Wilson Loops'', monopoles and so on \cite{pisarski,shuryak}).  
The scale at which such a degeneracy arises must be regulated via the scale at which Gribov copies dominate multiplied by the flow gradient, which however has no relation to the Knudsen number or interaction strength.
As we shall see, combining the symmetries of Lagrangian hydrodynamics \cite{shift} will non-Abelian gauge symmetry will allow us to see this ambiguity explicitly.
\section{The hydrostatic limit and non-Abelian gauge theory}
\subsection{Colorful swimming ghosts \label{swim}}
For a systematic look into this issue let us start from the first part of \cite{shift}.  There, it is shown that for a general theory with continuous media (three fields $\phi_I$, representing the lagrangian coordinates of the fluid cell) and internal conserved currents (The $\phi_I$ acquire a complex phase $\alpha$) ideal hydrodynamics is equivalent of imposing a Lagrangian depending not on $\phi_I,\alpha$ but on $b,y$.
\begin{equation}
  \label{syms}
F(\phi_I e^{i \alpha}) \rightarrow F(b,y) \eqcomma b=\left( \mathrm{Det}_{IJ}\left[ \partial_\mu \phi_I \partial^\mu \phi_J  \right] \right)^{1/2} \eqcomma y=J^\mu \partial_\mu \alpha
\end{equation}
where
\begin{equation}
\label{shift}
  J^\mu \propto u^\mu \eqcomma u_\mu = \frac{1}{6b} \epsilon_{IJK}
  \epsilon_{\alpha \beta \gamma \mu} \partial^\alpha \phi^I \partial^\beta \phi^J \partial^\gamma \phi^J
\end{equation}
and $F(...)$ is the lagrangian of the fluid, analogously to \cite{shift} it will be a Legendre transformation of the free energy w.r.t. the chemical potential (not to be confused with the Yang-Mills field, denoted in this paper by $G$). It is a simple exercise to show that Eqs \ref{syms} and \ref{shift} lead to Eq. \ref{hydrot} when Eqs \ref{tmunucan} and equivalently Eq. \ref{blderiv} are applied, underlining that this approach is equivalent to ``standard`` ideal hydrodynamics.

Here, the entropy $b$ dependence in Eq. \ref{syms} is equivalent to imposing invariance under all volume preserving diffeomorphisms and in addition Eq. \ref{shift} imposes invariance under $\alpha \rightarrow \alpha +f(\phi_I)$, the chemical shift symmetry\cite{shift}. Physically, chemical shift imposes the fact that any gradient of either chemical potential and density is proportional to velocity.   Mathematically, the phase in the internal symmetry becomes a function of $\phi_I$: Conservation laws ensure that any dynamics is a function of phase differences, and the gradient o the phase is exclusively in the $u_\mu$ direction.

Since Gauge symmetry is a symmetry in internal space, it is this symmetry that we will have to expand.  Let us therefore generalize
\begin{equation}
  \label{gauggen}
y= J^\mu \partial_\mu \alpha \rightarrow \left[ J^\mu \right]_a  \partial_\mu \left[ \alpha \right]_b = y_{ab} 
\end{equation}
to describe the color charge carried by the Lagrangian particle/volume element.

This generalization follows naturally from Eq. (14) of \cite{shift}.  As is explained there, this combination is the only combination which has both invariance under internal symmetries, the right order in derivative.  When the symmetry is more complicated than $U(1)$, with many generators and phases, current and phase can be in different directions.   To connect it to the more usual representation of chemical potentials, one must remember that for any charge in equilibrium $\mu_{\overline{Q}}=-\mu_Q$.     
For global symmetries such as Isospin or Flavor, superselection sectors (particle identities and Fermi surfaces) introduce a preferred basis in $ab$ where this chemical potential is defined \cite{piazza,braunerpiazza}.  In this case, for $N$ charges an $N\times N$ matrix with $N$ parameters
$y_{ab}=\mu_a-\mu_{b}$, diagonalized in the physical basis, would carry the full chemical information of the system.
 
Let us however consider color rather than flavor, and assume $y_{i}$ to be ``color charge'' chemical potentials, with no preferred basis, where rotations in color space yield a continuum of chemical potentials representing a continuum of conserved charges.   As previously shown in \cite{cfl,cflbla,cflbra,cflsed,cflred,cflba}  in the context of Color-Flavor-Locked matter, it means making $y_{ab}$ gauge {\em covariant} adjoint matrix, with as many chemical potentials as generators . While color charges are neither gauge invariant nor covariant, rotations in color space are gauge covariant and hence can be used to construct gauge invariant free energies \footnote{Strictly speaking one must connect every charge to $U_\infty(x)$, a Wilson line going to infinity to have well-defined charges up to topological ``large gauge`` transformations, see for example \cite{consti}.  Here, we neglect this, although it can be speculated the contradictions found in this section are related to this separation being impossible in the strongly coupled limit }.  Physically, such rotations are gauge-covariant color {\em currents}, which in local equilibrium are parametrized by rotation matrices.  This justifies the elevation of $J_\mu \partial^\mu \phi_I$ to a matrix parametrized by the full number of generators of the gauge group, the number of ''gluons''. In a straight-forward generalization of \cite{shift}, any combination of color-anticolor charge $J_{ab}$ will be 
\begin{equation}  \ave{J_{ab}^\mu} = \frac{dF}{dy_{ab}} u^\mu  
\label{currenteq}
\end{equation}
as a consequence of the chemical shift symmetry (note that in CFL matter, condensates break covariance \cite{cfl,cflbla} so Eq. \ref{currenteq} will give rise to a gap equation giving a preferred basis in color-flavor space.  This will not happen here).

Other approaches, such as Zubarev hydrodynamics, will yield the same result as discussed towards the end of the paper\footnote{Eq. \ref{rhoeq}, with both $T_{\mu \nu},n_\mu$ and $J_{\mu}$ becoming gauge dependent but the whole exponent gauge-invariant}.

Within a locally equilibrated fluid, such a chemical potential corresponds, in analogy to electromagnetism, to the effect on the fluid of the chromo-electric potential (a magnetic potential would break isotropy).
In initial conditions based on color domains \cite{domains}, such an initial color chemical potential would rapidly appear.

Let us therefore try to impose invariance under the gauge Symmetry.  
Indeed, the Yang-Mills Lagrangian is known to have a number of conserved currents that matches that of generators $T^a$
\begin{equation}
\label{conslag}
J^{\mu a} \rightarrow  D_\nu G^{\nu \mu a} + T^{a} J^{\mu a} \eqcomma D_\mu = -\partial_\mu + f_{abc} A^{b}_{\mu}... 
\end{equation}
What we need to do is to combine this definition with local isotropy, combining the definition in Eq. \ref{conslag} with local isotropy.
 
Throughout we shall assume a Lorentz covariant or a comoving gauge in order not to spoil isotropy explicitly
\begin{equation}
  \label{gaugedef}
F(y,...) = F(U^{-1}(x) y U(x)) \eqcomma U_{ab}(x) \in SU(N) = \exp \left[ \sum_i \alpha_i (x) \hat{T}_i  \right]
\end{equation}
At first sight, any term dependent on $|y_{ab}|^2$ will do.   One must remember,however, that ''color chemical potentials'' $y_{ab}$ do not have to be gauge-invariant, but they have to be gauge-covariant, to allow for the lagrangian to be gauge invariant.

Comparing Eq. \ref{gauggen} and Eq. \ref{gaugedef} one gets
\begin{equation}
  y_{ab} \rightarrow U^{-1}_{ac}(x) y_{cd} U_{bd}(x) = U^{-1}(x)_{ac} J^\mu_f U_{cf} U^{-1}_{fg}  \partial_\mu \alpha_{g} U_{bg} =
\end{equation}
\[\ =
  U^{-1}(x)_{ac} J^\mu_f U_{cf} \partial_\mu \left(  U^{-1}_{fg}  \alpha_d  U_{bd}(x) \right) - J^\mu_a \left( U \partial_\mu U \right)_{fb} \alpha_f
\]
the first term is automatically satisfied if $\alpha$ and $J$ transform in the fundamental representation under the gauge group.   The second term is impossible to satisfy without introducing additional degrees of freedom, represented by Gauge fields
\begin{equation}
  \label{neweos}
     F\left(b,J^\mu \partial_\mu \alpha \right) \rightarrow F\left( b, J^\mu \left( \partial_\mu - U(x) \partial_\mu U(x) \right) \alpha \right)
\end{equation}
Continuing in this direction and building a gauge field out of the $U(x)$ will give us a pure gauge classical theory of the type examined in \cite{jackiw,surowka}, where an ideal fluid interacts with a zero-entropy coherent classical field.

However, we would like to explore the opposite limit, where the field is in local equilibrium with the fluid, and its oscillations are dictated not by the classical equations of motion but by the maximization of free energy.
This is the motivation, in analogy with \cite{shift}, of defining the Lagrangian as a Legendre transform of the free energy.  This is of course not the only, and not the most popular, way to derive the hydrodynamic limit (transport theory and AdS/CFT are much more used within the relativistic context).   However, it is the only approach explicitly based on local entropy maximization, which makes it more suited to understand what happens when, as in gauge theories, microscopic correlations introduce a degeneracy in the minimum.

To incorporate gauge symmetry in this framework we would like to impose the chemical shift symmetry,
\begin{equation}
  \label{surprise}
 J^\mu_{ab} = \frac{\partial F}{\partial y_{ab}} u^\mu \eqcomma L= F(b,y_{ab} \left(1-u_\mu \partial^\mu \alpha^{i} ) \right)  \simeq F\left( b,Tr \left[ y_{ab} \left(1- (\hat{T}_{bc})_i u_\mu \partial^\mu \alpha^{i}  \right) \right]^2,... \right)
\end{equation}
The last term can be thought of as giving interactions between the different chemical potentials within the fluid.  Any infinitesimal change in $U$ is always $\delta U \sim \sum_i \delta \alpha_i \hat{T}_i$ where $\hat{T}$ are the generators.   Analogously to \cite{shift} local equilibrium ensures only $\delta \alpha_i$ in the direction of $u^\mu$ can change the dynamics.   The number of independent components of $y_{ab}$ in gauge space is indeed equal to the number of generators.

In electromagnetism
\begin{equation}
  \hat{T}_i \rightarrow 1 \eqcomma y_{ab} \rightarrow \mu_Q \eqcomma u_\mu \partial^\mu \alpha_i \rightarrow A_\tau
\end{equation}
i.e. $y_{ab}$ is just the charge chemical potential and the gauge remnant
corresponds to the
direction of the electromagnetic vector potential $A^\mu$ in the co-moving time $\tau$, can always be eliminated by a gauge choice (provided the electromagnetic field is locally equilibrated it is trivially zero, in that gauge choice also!), which we expect since, as explained in the introduction, electromagnetism is not expected to have dynamical degeneracies.   Note that in the opposite planar limit, if $y_{ab}$ has order unity $U_{ab}$ comes with $1/N$ factors, so planar expansions should miss the second term in Eq. \ref{surprise} at leading order.

Because of the ''twisting'' in color space of non-Abelian gauge theory, $\hat{T}_i u_\mu \partial^\mu \alpha_i$ can-not be reduced to a comoving time derivative.
This makes it appear a Gauge-invariant theory is in a sense never locally equilibrated, since $u_\mu$ must enter the Lagrangian even in the ideal hydrodynamic limit.  

To understand the physical consequences, and physical meaning of equations \ref{neweos} and \ref{surprise}, one would have to expand around the hydrostatic limit (where we fixed the gauge of the background chemical potential in the first direction)
\begin{equation}
  \label{soundfree}
\phi_I = X_I + \vec{\pi}_I^{sound} + \vec{\pi}_I^{vortex} \eqcomma \nabla.\vec{\pi}_I^{vortex}=\nabla\times   \vec{\pi}_I^{sound}=0 \eqcomma y_{ab}=\sum_i \left( \delta_{i1}+ \delta \alpha_i(x) \right)T_{ab}^i
\end{equation}
One can immediately see from Eq. \ref{surprise}  that  perturbations in $\delta \alpha_i$ will result in a negative free energy.  We know, from zero temperature non-Abelian field theory that negative-norm ``ghost'' states exist and have non-trivial dynamics.   The physical motivation for this is that they correspond to non-physical perturbations normalizeable into rotations in gauge space.  The negative norm is explained by the fact that these are ``negative paths'' which, when subtracted, remove gauge ambiguities from scattering amplitudes.
Since we have built our hydrodynamic lagrangian around the local maximization of entropy, it is natural that such negative norm states appear here, this time corresponding to ``negative microstates'', related to each other via gauge transformations.      Transverse fluctuations, corresponding to closed loops in configuration space, will also receive corrections from polarization but, as shown in the next subsection, these will not change the main issue highlighted here.

A more physical picture has been known for a long time within non-relativistic fluid dynamics:  A swimmer can \cite{swim1,swim2} move themselves with no net force in a time-reversible fluid (for the non-relativistic limit this is a compressible highly viscous fluid) because, at each second, they move within the ``gauge space of shapes'' allowable to their body.This class of problems is popularized by the famous ''falling cat problem'' \cite{cat}: a cat can always land on its feet despite not having anything to push against because, again, angular momentum conservation is not enough to ''fix the gauge''.
The ''colorful swimming ghost'' non-hydrodynamic modes derived here can be thought of as a field of such ``swimmers'', each in a gauge adjacent configuration (the ``arms and legs'' are in gauge space) and each within a neighboring fluid cell.  These modes will connect neighboring cells with no advective flow, something impossible in the usual Euler equation.

The problem is that the hydrostatic vacuum, seen in this form, becomes unstable.  For every entropy perturbation $\delta b$, one can, for a monotonic equation of state, solve Eq. \ref{neweos} and Eq. \ref{surprise} for corresponding $\delta \partial_\mu \alpha_i$ such that
\[\ \frac{\partial^2 F}{\partial_\mu \partial \pi_I} \partial_\mu \pi_I = \frac{\partial^2 F}{\partial_\nu \partial y_{ab}} \partial_\nu \alpha_i T_{ab}^i  \]
with a continuous ring in the $i$ direction being possible.

for such perturbations
\begin{equation}
\label{equalaction}
  F\left( b+\delta b,Tr \left[ y_{ab} \left(1- (\hat{T}_{bc})_i u_\mu \left( \partial^\mu \alpha^{i} + \delta \partial^\mu \alpha^{i} \right) \right) \right]^2 \right) =  F\left( b,Tr \left[ y_{ab} \left(1- (\hat{T}_{bc})_i u_\mu \partial^\mu \alpha^{i}  \right) \right]^2 \right)
\end{equation}
hence, a compression wave in entropy can {\em always} be cancelled out by a wave in chemical potential and the vacuum is unstable against these perturbations.
Furthermore, the directions of the gradients for this occurs generally are {\em not} parallel to $u_\mu$, and the assumption that all currents are proportional to $u_\mu$ is not physically realized.

The main result of this section is that our fluid, described with fundamental degrees of freedom, will be filled with non-hydrodynamic modes whose current will not be proportional to velocity.  It is a good confirmation of the intuition we developed in the introduction we speculated that the expansion in oriented Wilson loops will generally not commute with the expansion of the gradients in conserved quantities.

In this calculation we have considered only ideal hydrodynamic corrections to the hydrostatic limit.  However, the current formalism can be extended to also incorporate dissipative corrections (\cite{gtlagrangian} and references within) using the formalism of doubled variables.
In this formalism, all variables are doubled and two copies of the free energy are included in the Lagrangian
\[\  F(b,y,u_\mu) \rightarrow F(b_1,y_1,u_{\mu 1})-F(b_2,y_2,u_{\mu 2}) + G(b_1,b_2,y_1,y_2,u_{\mu 1},u_{\mu 2})  \]
$G(b_1,b_2,y_1,y_2,u_{\mu 1},u_{\mu 2})$ will contain both dissipative and anti-dissipative terms.
Choosing a future-pointing direction guarantees that the resulting equation of motion will be dissipative (the final state is not unique w.r.t. initial conditions) rather than anti-dissipative.   However, the fact that the hydrodynamic limit is unstable against color charge perturbations means that this instability will be contained in both copies of the free energy, in $F(b_1,y_1,u_{\mu 1})$ as well as $F(b_2,y_2,u_{\mu 2})$.   Hence, just like typically viscosity does not remove critical fluctuations in the vicinity of a critical point, dissipative corrections will not remove the fluctuations examined here.
\subsection{Colorful swirling ghosts \label{swirl}}
Since a term of the form $u^\mu \partial_\mu..$ appears in the Lagrangian, let us now investigate a situation where one of the currents experiences a non-zero vorticity.
We first note that, as discussed in \cite{nicolis1} vorticity conservation within Lagrangian hydrodynamics as a symmetry of the action under local volume-preserving diffeomorphisms.   Once polarization is added as an auxiliary field \cite{gt1,gt2,gt3} this diffeomorphism symmetry is broken but total angular momentum conservation is ensured by the Lagrangian's invariance under rotational symmetries, which could be spontaneously broken by local magnetization \cite{gt3}.  The exact division of angular momentum between microscopic polarization (called $y_{\mu \nu}$ in \cite{gt1}), its non-equilibrium generalization (called $Y_{\mu \nu}$ in \cite{gt3}) and vorticity is regulated by the equation of state \cite{gt1} and relaxation time \cite{gt3}.

What is new in Gauge theories is the appearance of a Gauge invariant term coupling a closed gauge current with gluon polarization
\begin{equation}
  \oint J_{i}^\mu dx_\mu   \equiv \int_\Sigma d \Sigma_{\mu \nu} \omega^{\mu \nu}  \ne 0 \rightarrow  \omega_{i}^{\mu \nu}= \epsilon^{\mu \nu \alpha \beta} \partial_\alpha J_{\beta ab} \ne 0
  \end{equation}
This is because the vorticity of a color current is not invariant under a gauge transformation, but it transforms in the same way as the Wilson loop.  In fact, the Wilson loop is nothing else but a vortex in gauge space rather than in flow space.  
\[\ \oint dx_\mu \partial^\mu U_{i} \equiv \int_\Sigma d \Sigma_{\mu \nu} (G^{\mu \nu}_{i})_i  \]
here $G^{\mu \nu}_i$ is the field strength, the Yang-mills generalization of the electromagnetic field, which is not gauge invariant.    
Thus, terms such as $Tr_{i} \left[ \omega_{\mu \nu} G^{\mu \nu} \right]$ can also enter the Lagrangian, and are at the same order as $u_\mu \partial^\mu \alpha$.

In \cite{jackiw}, these terms are interpreted as force terms in a Vlasov-type plasma.   Here, this enters the free energy so there is no force, it is a degree of freedom w.r.t. entropy is maximized.   It is therefore to be interpreted as the gluon polarization tensor, and such a term describes the ''chiral vortaic`` and ''chiral separation'' effects \cite{kharzeev}.

Hence, it is indeed true that a Gauge-invariant fluid is polarized.   However, we rather unexpectedly found, via Eq. \ref{surprise}, that its free energy, via the ``color chemical potentials'', must depend explicitly on velocity.
As a result, the polarization tensor, which in general has six independent components, is here determined by gauge structure, with $N^2-1$ redundant fields having 2 independent polarizations each.
Unlike the general polarization tensor $y_{\mu \nu}$ explored in \cite{gt1}, which has 6 degrees of freedom, here the equivalent is $N^2$ copies of $A_\mu^i$, which combine into $G^{\mu \nu}_i$ the usual way
\begin{equation}
G^{\mu \nu}_i = \partial^\mu A^\nu_i - \partial^\nu A^\mu_i + f_{i j k} A^\mu_j A^\nu_k
\end{equation}
The form of the equation of state in the small polarization limit should however be similar to that in \cite{gt1,gt2}, namely
\begin{equation}
  \label{eqlag}
  L= F(b,Tr_{i} \left[y_{ab} \left(1-u_\mu \hat{T}_{ab}^i \partial^\mu \alpha^{i} ) \right]^2,Tr_{i}  \left[ w_{i}^{\mu \nu} G_{\mu \nu}^i\right] \right) \simeq
\end{equation}
\[\    \simeq F(b\times \left[ 1-c\Omega^2+\order{\Omega^4}  \right],Tr_{i} \left[y_{ab i} \left(1-u_\mu \hat{T}_{bc i} \partial^\mu \alpha^{i} \right) \right] \eqcomma \Omega^2 \simeq \sum_i \left( G^{\mu \nu}_i \omega_{\mu \nu i} \right)^2
\]
This equation of state includes both swimming and swirling ghosts.

In equilibrium, polarization and vorticity must point in the same direction according to the arguments made in \cite{gt1}.  However, the local equilibrium limit is unstable as shown in \cite{gt3}.  The resulting relaxation dynamics will be affected by this different number of effective polarization degrees of freedom.
Following \cite{gt3}, $G^{\mu \nu}_i$ relaxes to $\omega^{\mu \nu}_i$ using an Israel-Stewart type equation.  The naive equivalent is
\begin{equation}
\label{relax0}
\tau u^\beta \partial_\beta \mathcal{G}^i_{\mu \nu} + \mathcal{G}^i_{\mu \nu} = \chi \omega^i_{\mu \nu} + \order{f_{ijk} \omega_j \omega_k}
\end{equation}
where
\[\  \chi \equiv \left| dF/d  G^i_{\mu \nu} \right|  \]
and $\mathcal{G}_{\mu \nu i}$ are non-equilibrium polarization fields (independent degrees of freedom with purely relaxational dynamics) and $G_{\mu \nu i}$ are the equilibrium values (determined by minimizing the free energy w.r.t. the local angular momentum and chemical potentials).

This equation can be obtained, using the doubled variable techniques, from a lagrangian of the type of Eq. (9) of \cite{gt3}, with no Gauge symmetry this lagrangian would read as
\begin{equation}
  \label{genlag}
L = F(...) + \mathcal{L}_{IS - vortex}
\end{equation}
where $F(...)$ is the equilibrium lagrangian of \ref{eqlag}  and $\mathcal{L}_{IS - vortex}$ contains  non-equilibrium mode for the Polarization tensor having a purely relaxational dynamics, equivalent to Israel-Steward dynamics developed in \cite{gtlagrangian}, with the doubling of the spin fields $\mathcal{G}^{\mu \nu i} \rightarrow \mathcal{G}_{\pm}^{\mu \nu i} $ taking care of the dissipative terms in the lagrangian
\[\ \mathcal{L}_{IS - vortex} = \frac{1}{2} \tau_Y \sum_i  \left( \,\mathcal{G}^{\mu\nu}_{i-} \, u^{\alpha}_{+} \partial_\alpha \mathcal{G}_{\mu\nu i +}
- \mathcal{G}^{\mu\nu}_{i+} u^{\alpha}_{i-} \partial_\alpha \mathcal{G}_{\mu\nu -} \right) +
\]
\begin{equation}
  \label{lagrangiang}
+ \frac{1}{2} \sum_i  \mathcal{G}_{i \mu\nu \pm}  \mathcal{G}^{\mu\nu}_{i \pm}+ \mathcal{G}^{\mu\nu}_{i \pm} F\left( \chi(b,\Omega^2) \omega_{\mu\nu} \right)
\end{equation}
Had Gauge invariance not been an issue, as illustrated in \cite{gt3}, this Lagrangian choice would have been uniquely determined by causality and the existence of a well-defined equilibrium state.   Polarization is a dynamical variable, but all it does is relax to its equilibrium value.   Hence, entropy is always close to the local maximum and the finite relaxation time assures causality.

However, a lagrangian such as \ref{lagrangiang} breaks Gauge invariance, since $\mathcal{G_+}$ and $\mathcal{G_-}$ are independent variables, and indeed the resulting equation of motion Eq. \ref{relax0} is not invariant under time-dependent gauge transformations ($\alpha$s in Eq. \ref{gaugedef} depending on $\tau$).
The only way to adjust the Lagrangian is to increase the powers of the $\mathcal{G}_{i \mu\nu \pm}  \mathcal{G}^{\mu\nu}_{i \pm}$ terms in the lagrangian to at least a power of $4$, with two $\mathcal{G}_+$ and two $\mathcal{G}_-$ terms.

However, this would inevitably preclude a unique relaxation minimum, since ``swirling'' solutions which rotate in gauge space and in configuration space at the same frequency
\begin{equation}
\mathcal{G}^{\mu \nu}_i \propto U^{i j}(x^\mu) \omega^{\mu \nu}_j(x^\mu) \eqcomma U^{ij} = \exp \left[ \sum_i \alpha_i(x) \hat{T}_i \right] \eqcomma \nabla_X \wedge \alpha_i \ne 0
  \end{equation}
will never relax to a value parallel to $\omega$, 
since a unique relaxation breaks gauge symmetry.   One can think of such solutions as being the vortex equivalent of the swimming ghosts and indeed, to leading order in gradient, they are captured in transverse modes of Eq. \ref{soundfree}.   What this section shows is that polarization corrections, at higher order in gradient, is prevented by Gauge invariance from relaxing vortical perturbations.

The fact that such vortices do not relax is not too surprising, since they should show up easily in the Wilson loop expansion \cite{pisarski}.   Because the Wilson loop and the vortex couple directly, a $Tr[U_a \omega_a]$ interaction appears at the lowest gradient level in both the Wilson loop expansion and the vorticity one.   This interaction has a direction in gauge space, an axial direction in configuration space, and, just like a hydrodynamic vortex, no propagating more or energy gap.   It is thus not surprising that it gives rise to a non-dissipative excitation mixing microscopic and collective degrees of freedom.  

Note that everything discussed in this section is related to {\em gluon} spin, as quark spin being gauge independent and its lagrangian can be constructed with no ambiguities \cite{gt1,gt2,gt3}.   As the proton spin puzzle shows, however, even with quark flavor a big percentage of spin degrees of freedom are concentrated within gluon matter, whose decomposition at the microscopic level \cite{araujo,goldman} is subject to the ambiguities illustrated in section \ref{tmunu}.   What this section shows is that, unlike quark spin, cannot achieve local equilibrium with vorticity.   This is analogous to the ambiguity of the hydrostatic limit w.r.t. sound waves.
\section{\label{discussion} Discussion}
In the previous sections, we have shown that, even if the thermalization scale is parametrically smaller than the color domain scale, if the color current is zero the hydrostatic local equilibrium limit will be plagued by gapless negative entropy ``excitations'' , reflecting the fact microstates within a non-Abelian gauge theory cannot be locally defined.   This, such a ``colored fluid'' will have a very different behavior from an ``ideal fluid''.

Our method, relying on bottom-up effective field theory via a Lagrangian construction developed in \cite{shift} is of course not the only, or the most popular, way to define hydrodynamics as a physical theory.   Indeed, hydrodynamics is usually treated in terms of {\em top-down microscopic dynamics}  around local equilibrium \cite{kodama}, either via transport or via quantum field theory constructions such as Schwinger-Keldysh and, recently, holography\cite{spal}.   In particular, one could ask to what extent is any result developed here unique to our approach rather than hydrodynamics in general.    The advantage of the approach developed by \cite{shift} is that local equilibrium is treated as an assumption, and the lagrangian is developed from this assumption.  By contrast, transport usually requires assumptions independent from the smallness of the Knudsen number, such as molecular chaos and the planar limit.   Thus, the ambiguities we have derived will show up in the most straight-forward way in the hydrodynamics defined via \cite{shift}.    We also reiterate that this ambiguity is explicitly dependent on local equilibrium, which, although is superficially similar to the ``hydrodynamization'' inherent in \cite{wong,heinz,jackiw,surowka,mrow,whitening}, is in fact generally very far away from the regime where approaches based on strong field transport apply.

To physically interpret the effects calculated in this work more generally than within our formalism one must understand how the Gauge ambiguities affect hydrodynamics defined as a limit of transport.  To understand this, one must remember
that a fluid has three scales, which are sequentially coarse-grained.
   Quantitatively, probability of thermal fluctuations is normalized by the heat capacity and temperature scale $1/(c_V T)$  and microscopic correlations due to viscosity are $\sim \eta/(Ts)$.  Since for a usual fluid, there is a hierarchy between microscopic scale, Knudsen number and gradient
\begin{equation}
\label{hyerarchy}
  \frac{1}{C_V T} \ll \frac{\eta}{(Ts)} \ll (\partial u_\mu)^{-1}
\end{equation}
The first inequality defines the truncation of microscopic correlations within transport (As long as the first inequality holds microscopic fluctuations and correlations will be dissipated sooner than any hydrodynamic response) and it also automatically holds in the planar limit in the gauge/gravity correspondence.
It is this inequality which guarantees that the complexities of hydrodynamics (the existence of ''wild'' solutions \cite{wild}, turbulence, etc.) can coexist with a well defined statistical mechanics applicable to the hydrostatic limit.
The second inequality is usually associated with the Knudsen number, it avoids microscopic  correlations between different fluid cells and hence allows for an expansion in either gradients of conserved quantities or moments of the microscopic distribution function \cite{kodama}.  

Comparing Equation \ref{hyerarchy} with Equation \ref{surprise} it is clear that the ``smallest scale`` and the ``largest scale`` in Eq. \ref{hyerarchy} will always be of the same order, since perturbations in $\delta y_{ab}$ (charge density waves) and $\delta u_\mu \alpha_i$ (swimming ghosts) are related by a factor of unity.   The left-most term in Equation \ref{hyerarchy} is expected to be of order of the distance between microscopic degrees of freedom, while the right-most term is a macroscopic sound wave, but Gauge invariance introduces redundancies independent of frequency.   
Of course, if the microscopic theory is asymptotically free, for higher frequencies the assumption of local equilibrium under which Equation \ref{surprise} will be less and less tenable ($\eta/(Ts)$, or, more accurately, the color diffusion scale will dissipate gradients in $y_{ab}$ ``instantaneusly``), but as long as the the characteristic scale of sound waves is in the strongly coupled regime at the hydrodynamic scale gauge symmetry breaks any scale separation between sound waves and microscopic motion.

Hence, a colored fluid close to ideal equilibrium cannot be reduced to the type of ''interacting quasi-particle picture'' for which a Boltzmann or a Boltzmann-Vlasov equation are appropriate.   While one imagines perturbative gluons to be modeled via a Boltzmann equation \cite{cascade}, and the effect of coherent fields to reduce to an ''anomalous viscosity'' \cite{asakawa} when the direction of the fields is random enough, ghost fields cannot be pictured this way precisely because they have a ''negative'' effect on the fluid at the level of the density matrix \cite{densitymatrix}
\begin{equation}
  \label{densitymatrix}
  \hat{\rho}=\mathcal{Z}^{-1} \int \mathcal{D}\phi <\phi|\Psi><\Psi|\phi> \underbrace{\rightarrow}_{Boltzmann} f(x,p) \delta \left( <\phi|\Psi> - f(x,p)  \right)
\end{equation}
, which can-not factorize into microscopic particle distribution functions (the ``$\underbrace{\rightarrow}_{Boltzmann}``$ step ) even approximately.    Just like, in QCD, the effective action $\lnz $ is not even close to the semiclassical expectation value $\mathcal{Z} \simeq \exp\left[iS\right]$ because quantum fluctuations change the vacuum, the free energy of the locally equilibrated fluid will have little relation to the classical equation of state.      Physically, equation  \ref{soundfree} and \ref{equalaction} demonstrates that separating microscopic changes in the distribution of physical degrees of freedom (where entropy changes within the cell) from ``Gauge`` degrees of freedom (which are part of the same microstate) is complicated by the non-linearities of the theory, but this is a qualitative indication that at a momentum scale where sound-waves appear, only color neutral fluctuations are part of the physical spectrum of excitations, analogously to how only color neutral light excitations appear once a Gribov horizon is imposed in quantum field theory \cite{marcelo0,marcelo}.

In fact, our approach can be rewritten using the Zubarev formalism
\cite{zubarev,becfluid}, where the density matrix of an evolving fluid in local equilibrium is written in terms of infinitely many Lagrange multiplies each describing temperature, velocity and chemical potential fields
\begin{equation}
\label{rhoeq}
\hat{\rho} = \frac{1}{\mathcal{Z}(T(x),u_\mu(x),\mu(X))}  \exp  \left[ - \int_\Sigma d\Sigma(\tau)  \frac{u_\mu \hat{T}^{\mu \nu} \hat{j}_\nu - \hat{j}^\mu n_\mu}{T}  \right]
\end{equation}
under the local minimization ($u_\mu,T,\mu$ are defined at each point in space) of the entropy operator 
\[\ s= \mathrm{Tr} \hat{\rho} \ln \hat{\rho}\]   
It is a short and straight-forward derivation \cite{becfluid} to show that for a minimum to be well-defined in Eq. \ref{rhoeq} one must have $u_\mu$ and $n_\mu$ parallel to killing vectors of $\Sigma$, the equivalent of Eq. \ref{shift}.   Conservation of $s u^\mu$ in the leading order expansion quickly follows, fully establishing the equivalence of the two approaches.
What we have shown in the paper, in the language of \cite{becfluid}, is that once $j^\mu$ and $T_{\mu \nu}$ become gauge-covariant, so must $u_\mu$ and $n_\mu$. i.e., tracing out gauge degrees of freedom breaks the uniqueness of the choice of $\Sigma$ and hence local thermal equilibrium.   In this formalism it is actually easy to see this is the case, since $n_\mu$ is defined via \cite{becfluid} a totally vorticity-free field
\[\ \epsilon_{\mu \nu \rho \sigma} n^{\nu} \left( \partial^\rho n^\sigma- \partial^\sigma n^\rho \right) =0 \]
However, conserved currents $j^\nu$ are gauge covariant, so any such closed loop transforms as a Wilson loop.

One obvious question to ask is, what if instead of choosing a Gauge we formulate our locally equilibrated theory  in a gauge-invariant way, via Wilson loops \cite{pisarski} where in principle microstates can be counted without resorting to ghosts.
Constructing an explicitly gauge-invariant theory compatible based around local equilibrium of Wilson loops is highly non-trivial, since a Lagrangian based on Wilson loops forms an infinite series and Wilson loops are non-local objects having an orientation in space.  This defines a gradient expansion (characteristic Loop size $\times$ gradient) unrelated to the Knudsen number.   
The limit where the series expansion and the gradient expansion ``commute'', so that microscopic physics to zeroth order is isotropic, could well be not realized, even for theories where the Knudsen number vanishes   Physically, as we know from, for example, polymer fluids and other soft condensed matter systems, in the strong-interacting regime one can get not an ideal fluid limit nor a transport regime but extended ``polymer-like`` orientable structures correlating macroscopic distances \cite{shuryak}.  It is far from clear local equilibrium in such a situation corresponds to an ideal fluid limit.   Thus, Wilson loops offer a physical picture that is compatible to the picture our calculation motivated:  Gauge ambiguities preclude a well-defined local equilibrium state or a transport regime when color charges are present because the two gradients (the flow gradients entering $T_{\mu \nu}$ via the Knudsen number and the Wilson loop gradients on which the equation of state depends) can not be disentangled.

Another obvious question is, why is there no trace of such non-hydrodynamic modes in holography, where everything converges to a Knudsen-based gradient expansion with a usual equation of state.   The answer is that gauge/gravity duality, as done so far, requires a planar limit and a conformally invariant ultraviolet fixed point.  Color flying ghosts, and the difference between the Belinfante and Canonical tensors are of order $\order{N}$, just like thermal perturbations, while thermal degrees of freedom are of $\order{N^2}$.  Hence, they do not contribute to the planar limit.   The non-local Gribov copies continue \cite{marcelo} continue to exist but because the number of gluon microstates is parametrically larger, the planar limit does not see their contribution.

One could still ask why these deviations are however not present in calculations where corrections to the planar limit are manifest \cite{petrov}.
A second possible issue is conformal invariance, which is known to restrict ghost modes in field theories.  Conformal symmetry fixes pseudo-gauge transformations of Eq. \ref{pseudogauge} to a form  determined up to a scalar field $\phi$ \cite{fabbri}
\begin{equation}
  \label{cftpseugo}
  \Phi_{\lambda,\mu \nu} \underbrace{\rightarrow}_{conformal} g_{\sigma \mu} \partial_\nu \phi - g_{\sigma \mu} \partial_\mu \phi
  \end{equation}
which makes no effect on any component of the energy-momentum tensor.
A $T_{\mu \nu}$ in local equilibrium with a fixed gauge should thus be unaffected by ghosts.
Of course, as we have seen $T_{\mu \nu}$ and its conservation should not uniquely determine the dynamics \cite{gt1}, and, in any case, in
Generic gauge/gravity constructions have conformal symmetry only as a renormalization group limit.
However, as shown in \cite{marcelo}, a conformally invariant fixed point makes the Gribov issue microscopically non-dynamical, leading to the suspicion it will not contribute to any locally equilibrated dynamics either.  Certainly, a gradient expansion in terms of either ghosts or Polyakov loops seems forbidden by conformal symmetry.     Ghosts continue to exist, but conformal symmetry forbids ''second Knudsen numbers'' dependent only on microscopic fluctuations.
An exploration of weather this makes non-hydrodynamic ``ghost'' modes be irrelevant requires developing a linearization of the modes in Section \ref{swim} and \ref{swirl} for currents obeying the algebra of $\mathcal{N}=4$ super-Yang mills, something best left to a follow-up work once a linearization of the theory based on swimming and swirling ghosts is completed.

What is the role of the swimming and swirling ghosts in the dynamics of a close-to-ideal fluids in non-Abelian gauge theory?   A linearization and causality analysis of this system is left for a forthcoming work.  We note, however, that as in \cite{gt1,gt2,gt3} Ostrogradski's theorem means that such non-hydrodynamic modes usually generate instabilities and causality violation.  The only way to make such modes go away is to insure local color neutrality (zero chemical potential everywhere in the system), leading to the suspicion that these non-hydrodynamic modes quickly color-neutralize and locally thermalize the system.  A qualitative manifestation of the dynamics described here is that the fluid created in heavy ion collisions is color neutral on scales parametrically smaller than the mean free path once thermalization occurs.   The idea of an undefined local equilibrium for a color charge can be thought of the local thermal equivalent of the criterion of confinement that ``a colored state never goes on-shell`` associated with infrared positivity violation \cite{alkofer} scenarios.   
The qualitative picture of confinement behind such a scenario is that a colored particle chaotically radiates soft virtual gluons until color-neutralization.    Turbulent non-hydrodynamic modes are a way to implement this in a locally thermalized medium.   A similar conclusion was actually reached a long time ago \cite{whitening}, but there the color-neutralization scale,while being significantly shorter than other scales, was dissipative.   Here we show this scale is more similar to the stochastic scale governing thermal fluctuations, driven by the microscopic degeneracy rather than the mean free path, and generating fluctuations rather than dissipation.  

One could also speculate that ghosts might have something to do with the fact that collectivity in hadronic collisions seems to be independent of the number of degrees of freedom \cite{smallflow}, something naively at odds with the hydrodynamic picture since fluctuations should be inversely proportional to the degrees of freedom available to the system \cite{vogel}.   Ghosts could give a source of ''negative fluctuations'' that bring the system closer to the equilibrium state, and scale in the same way as the usual thermal fluctuations.   A lattice implementation of colored hydrodynamics, achievable with the methods of \cite{burch}, would quantitatively investigate this.

In conclusion, we find that the macroscopic symmetries of ideal hydrodynamics are generally incompatible with the microscopic symmetries non-Abelian gauge theory.  The ideal fluid limit of a theory whose microscopic dynamics has such a symmetry, therefore, is very different from the Euler equations, as it will be full of non-hydrodynamic ``ghost`` modes carrying rotations of color space along the flow direction.  The only fluid dynamic limit where something like an Euler equation, with an equation of state independent of flow emerges, is one where color neutrality is assured in each volume cell.  The consequences of this for quark gluon plasma thermalization is likely to be profound.

\textit{Acknowledgements}   GT acknowledges support from FAPESP proc. 2017/06508-7,
partecipation in FAPESP tematico 2017/05685-2 and CNPQ bolsa de
 produtividade 301996/2014-8. DM was supported by CNPQ graduate fellowship n. 147435/2014-5. 
This work is a part
of the project INCT-FNA Proc. No. 464898/2014-5.
We wish to thank David Montenegro, Radoslaw Ryblewski, Saso Grozdanov, Luca Fabbri and Henrique Sa Earp for fruitful discussions.


\begin{thebibliography}{99}
\bibitem{jackson} for a worked out example, see\\
  J.D.Jackson, ''classical electrodynamics'', third edition, Chapter 12.10\\
  Of course there the gauge theory is electromagnetism, where photons never reach local equilibrium, so local equilibration in relation to the regularisation of $T_{\mu \nu}$ is not discussed
\end{thebibliography}
\end{document}